\documentstyle[aps,multicol,epsfig]{revtex}
\draft

\begin{document}
\author{Fr\'ed\'eric Lacombe$^1$, Stefano Zapperi$^{1,2}$ and Hans J. Herrmann$^1$}
\title{Force fluctuation in a driven elastic chain}
\address{$^1$ PMMH-ESPCI, 10 rue Vauquelin, 75231 Paris Cedex 05, France}
\address{$^2$ INFM, Dipartimento di Fisica, E. Fermi, 
        Universit\`a "La Sapienza", P.le A. Moro 2,
        00185 Roma, Italy }
\maketitle

\begin{abstract} 
We study the dynamics of an elastic chain driven on a disordered substrate
and analyze numerically the statistics of force fluctuations at the depinning
transition. The probability distribution function of the
amplitude of the slip events for small velocities
is a power law  with an exponent $\tau$ depending
on the driving velocity. This result is in
qualitative agreement with experimental measurements
performed on sliding elastic surfaces with 
macroscopic asperities. We explore the properties of the 
depinning transition as a function of the driving mode 
(i.e. constant force or constant velocity) and
compute the force-velocity diagram using finite size scaling methods.
The scaling exponents are in excellent agreement with the
values expected in interface models and,  contrary to previous 
studies, we found no difference in the exponents for periodic and  
disordered chains. 
\end{abstract}
\begin{multicols}{2}
\section{Introduction}    \label{sec:introduction}
The dissipative motion of an elastic line in a random potential is 
an interesting example of a nonequilbrium interacting system
and is relevant for several phenomena in condensed matter physics.
Examples are numerous and include the motion of magnetic 
interfaces in ferromagnetic materials \cite{Zapperi1,Zapperi2}, 
solid friction \cite{persson,Hwa,Bocquet}, wetting \cite{ERT-94,Raphael},
charge density waves \cite{CDW}, 
fluids in porous media \cite{robbins}, vortex dynamics in high 
temperature superconductors \cite{Ertas,Zapperi3,Ledoussal}, cracks  and
dislocations. These systems are characterized by 
a dynamic phase transition ruled by
the interplay between quenched disorder 
and elastic interactions. 

Due to the effect of the disorder, 
an elastic chain at zero temperature is pinned when the applied force is 
below a critical value $F_c$:
after a sufficiently long time, independently of the initial
conditions, the chain reaches a configuration
where no movement is possible. For $F>F_c$ the chain can escape from
any pinning configuration and moves
with constant average velocity.  When $F$ is close to $F_c$ 
the motion is dominated by collective effects and 
the depinning of a single bead produces a large reorganization of the 
chain. In other words, for $F=F_c$ the system  
is {\em critical} and the motion of the beads is highly correlated.  

An elastic chain moving in a disordered potential 
is a useful model to understand some general features
of sliding friction \cite{Bocquet} in particular of 
the experiment reported in Ref.~\cite{Laroche,Laroche2,Laroche3}, 
done using  two artificial surfaces with controlled roughness and elasticity.
Beads of diameter 2mm  were randomly put inside an elastic matrix, 
with a maximum roughness of 0.5mm. 
The two surfaces were then displaced against each other
at constant velocity and the friction force was measured, varying
the elasticity of the matrix and the driving velocity. 
The distribution of the amplitude of the slip events 
is generally found to decay as a power law at
small velocities, suggesting the presence of an underlying critical point. 
The  exponents characterizing the power law distribution 
are found to decrease with the applied velocity, in analogy
with other driven systems such as domain walls in 
ferromagnet \cite{Zapperi1}.

Several variant of the chain model can be studied in order to
reproduce the experiments: periodic or
disordered \cite{Hwa} arrangements of beads, on a rigid \cite{Hwa,Bocquet}
or elastic \cite{Matsukawa2} substrate.
Simulations of a one dimensional, periodic or disordered, chain over 
a rigid disordered potential have been performed by Cule and Hwa \cite{Hwa}. 
The measurement of the velocity and the roughness exponents seem
to indicate that periodic and disordered chains are described by two 
different universality classes. These and other simulations are performed
considering a constant applied force, while experiments are performed
driving the system at constant velocity. 

Here we consider explicitly the second case and 
measure the force fluctuations as a function
of the applied velocity. The distribution of the slips events 
is power law distributed,  and characterized
by an exponent $\tau$, which appears to decrease with the 
applied velocity, in agreement with the experiments. 
In the limit of low velocity the exponent $\tau$ can be related
to the critical exponent obtained tuning the applied force 
as first discussed in Ref~\cite{tang-bak} 
(see also \cite{Zapperi1,Hwa,Fisher3,maslov}).
We measure the fluctuation in the position of the beads in
the constant force and constant velocity cases and obtain 
the same roughness exponent $\zeta$. Next we evaluate the
force velocity diagram, using finite size scaling to 
locate the critical force and compute the exponent $\beta$. 
The values of the exponents are consistent with scaling relation and
in good numerical agreement with the exponents of
interface depinning, but disagree with previous simulations
for a periodic chain in a random potential \cite{Hwa}.
In order to confirm this conclusion, we simulate the motion of 
a disordered chain and see no evidence for the existence of two 
different universality classes for periodic and disordered chains, 
in disagreement with the conclusions of Ref.~\cite{Hwa}. 

The paper is organized as follows in Sec.~\ref{sec:2} we introduce the
model, and in Sec~\ref{sec:2b} we 
define the critical exponents and discuss some scaling
relations. In Sec.~\ref{sec:3} we present the numerical
results obtained at constant velocity, 
in Sec.~\ref{sec:4} we discuss the constant force case
studying the scaling behavior close to the depinning transition. 
In Sec.~\ref{sec:conclusion} we summarize the main results of the paper.

\section{Model} \label{sec:2}

We consider the overdamped dynamics of a one dimensional line of 
elastically coupled beads, driven on a disordered substrate. 
The disordered potential is a succession 
of identical Gaussian potentials, randomly distributed in space. 
The beads can be driven directly, applying a constant force,  or
indirectly coupling them to an intermediate spring which is pulled
at constant velocity.
The equation of motion is, 
\begin{equation}
\eta \frac{\partial r_i(t)}{\partial t} = D (r_{i+1}- 2 r_i + r_{i-1}) +
f(r_i(t)) +F,
\end{equation} 
where $r_i(t)$ is  the position of the bead $i$ at time $t$, 
$\eta$ is the coefficient of viscosity,  
$D$ is the stiffness of the elastic line, $F$ is the driving force, and
$f(x)$ is a random force, due to the contribution of 
the ensemble of pinning centers. We model the random force by the sum of
$N$ derivatives of Gaussian potentials located on the pinning site
\begin{equation}
f(x) = C \sum_{i=1}^N  (x-x_i^p)
\exp\left[-\frac{1}{2}\frac{(x-x_i^p)^2}{\sigma^2}\right],
\end{equation} 
where $C$ represents the strength of the disorder,
$\sigma$  quantifies the width of the wells, and   
$x_i^p$ is the location of the pinning site $i$, 
which we  chose to be Poisson distributed.
The equation of motion is integrated numerically using a fourth-order
Runge-Kutta method.

The interplay between disorder and elastic interactions in our model
can be understood computing the Larkin length $l_L$ \cite{Larkin}.
For distances smaller than $l_L$, 
the beads are interacting strongly and the chain moves coherently,
while for distances larger than $l_L$ the random forces become dominant
and the chain deforms considerably. 
The Larkin length can be estimated considering 
both the effect of the rigidity of the 
line and the strength of the disorder and for our model it is
given by  \cite{Bocquet}
\begin{equation}
l_L \approx [ \frac {D a \sigma^2}{\rho^{1/2} C } ] ^{2/3},
\end{equation}
where $a$ is the distance between the beads and $\rho$ is
the density of pinning centers.
In order to analyze the critical properties of the system,
we have to consider the limit where the Larkin length is 
larger than the mean distance between the beads and the dynamics 
is governed by the  collective  motion of the beads. To this 
end we carefully choose the parameters of the model so that
$l_L \gg a$.

\section{Scaling relations}\label{sec:2b}

Depending on the method used to drive the chain the measured quantities  
change, but the corresponding critical exponents can be related
by scaling relations. 
Here we summarize the scaling properties of the depinning transition
in the case where the beads are driven by springs of stiffness $K$ pulled
at constant velocity $V$ and in the case where they are submitted
to a constant force $F$.
 
\subsection{Constant velocity driving}

Friction experiments are generally performed under a constant velocity
driving. In some cases, a traction machine turning at constant velocity
is coupled by a spring to the sliding system. In other cases, an
effective spring coupling is due to the elastic deformation of the material
driven imposing a constant strain rate far from the sliding interface.
To simulate constant velocity driving \cite{note},  we attach each bead to 
a spring of stiffness $K$ \cite{Hwa}, so that the force is given by 
\begin{equation}
F = K (r_i(0)+V t  -  r_i(t))
\end{equation}
where $V$ is the applied velocity.
When the velocity $V$ and the stiffness $K$ are small, the 
motion displays large fluctuations. In particular, in the 
limit $V\to 0$ and $K \to 0$ the system reaches the
depinning transition and the force fluctuates around $F_c$.
This feature is common to other non equilibrium critical
phenomena, such as absorbing state phase transitions 
and self-organized criticality \cite{btw,Rev}.

The distribution of the friction force fluctuations is directly related to 
the size distribution of slip events $\Delta x$ of the chain, 
since $\Delta F = K \Delta x$. Here $x\equiv \sum_i r_i$ and
the slip is defined by a drop in the measured friction force
(See Fig.~1). Close to the depinning transition, 
we expect that the distribution
of $\Delta x$ decays as
\begin{equation} \label{eq:scaling1}
P(\Delta x) \sim s^{-\tau} h(\Delta x/\Delta x_0),
\end{equation} 
where $h(x)$ is a scaling function and $\Delta x_0$ the cutoff value.
The value of the cutoff depends on various parameters, such as the 
system size employed in the simulation. Clearly the avalanche 
size cannot be  greater than the total length of the line. 
Furthermore, we expect that the stiffness of the springs $K$ 
and the driving velocity $V$ will in general 
change the value of the cutoff. In the low velocity limit and 
for large enough system sizes, $K$ becomes 
the dominant parameter that determines the value of the cutoff \cite{Hwa}.

The scaling of the cutoff with $K$ can be used to evaluate
the roughness exponent $\zeta$, using the relation
\begin{equation} \label{eq:cutoff} 
\Delta x_0 \sim K^{-\zeta/2}.
\end{equation}
This relation  can be obtained noting that
the chain  can not jump over a distance larger than $\xi_0$,
which represents the length for which the elastic interaction term 
is overcame by the restoring force due to the springs:
 $D \xi_0^{-2} \sim K$. Then, using the scaling 
relation for $\Delta x_0 \sim \xi_0^{\zeta}$,
we obtain Eq~(\ref{eq:cutoff}) \cite{Hwa,Zapperi2}. 
This result implies that $P(\Delta x,K)$ should satisfy the scaling form
\begin{equation} \label{eq:scaling2}
P(\Delta x,K) K^{-\tau \zeta/2} = H(t),~~t\equiv\Delta x K^{\zeta/2}. 
\end{equation}
Eq.~(\ref{eq:scaling1}) is used to compute $\tau$  and 
Eq.~(\ref{eq:cutoff}) and Eq.~(\ref{eq:scaling2}) are used to compute 
the exponent $\zeta$.

The exponent $\zeta$ can also be evaluated directly
using the scaling of the fluctuations of the displacements
with $K$.  Defining the relative displacements of the beads 
as $u_i(t)\equiv r_i(t)-(Vt+r_i(0))$, 
the fluctuations can be quantified by
\begin{equation}
W= \sum_{i=0}^L (u_i(t) - m_i(t))^2 /L,
\end{equation}
where $m_i(t) \equiv \sum_{i=0}^L u_i(t)/L$, and $L$ is the number 
of beads. The roughness $W$ scales with the correlation length $\xi_0$ 
as $W \sim \xi_0^{\zeta}$, and since $\xi_0 \sim K^{-1/2}$
\begin{equation} \label{eq:scaling_rugo}
W \sim K^{-\zeta / 2}.
\end{equation}
Eq.~(\ref{eq:cutoff}) and Eq.~(\ref{eq:scaling_rugo}) have the 
same origin and can independently be used to estimate $\zeta$. 

\subsection{Constant force driving}

When the system is driven at constant force we expect a depinning
transition as a function of $F$. For $F>F_c$, the chain moves
with constant average velocity $v$ defining an exponent $\beta$
\begin{equation} \label{eq:scaling3}
v \sim (F-F_c)^{\beta}.
\end{equation} 
Close to the depinning transition the motion is very irregular,
and large regions of the chain move collectively. The correlation
length diverges at the transition as
\begin{equation} \label{eq:scaling3b}
\xi \sim (F-F_c)^{-\nu}.
\end{equation} 
In order to estimate the critical exponents $\beta$ and $\nu$ 
we employ a particular finite size scaling 
method \cite{Zapperi3}, in analogy with absorbing state 
phase transitions \cite{Dickman}. 

We first compute the critical force analyzing the decay of
the average velocity with time for different system sizes.
For finite systems, the average velocity reaches a quasi
steady state $v(F,L)$. When $F > F_c$ we expect that
as the size $L\to\infty$, $v(F,L)$ approaches a non vanishing value
given by Eq.~(\ref{eq:scaling3}), while it decays to
zero for $F < F_c$. At the depinning transition 
we expect that
\begin{equation} \label{eq:scaling4}
v(F_c,L) \sim L^{-\beta/\nu}.
\end{equation}
Once $F_c$ is known with good precision, we can measure
directly the exponent $\beta$ from  Eq.~(\ref{eq:scaling3}).
As in the constant velocity case, we can evaluate the 
roughness exponent, measuring the width at $F_c$ as a function 
of $L$, which should scale as
\begin{equation} \label{eq:scaling5}
W(L) \sim L^{\zeta}.
\end{equation}
For a periodic chain Ref.~\cite{Hwa} reported $\beta\simeq 0.4$
and $\zeta\simeq 1.5$, while for a disordered  
$\beta\simeq 0.25$ and $\zeta\simeq 1.2$. These last value
are consistent with interface depinning that in $d=1$ yields
$\beta\simeq 0.25$ and $\zeta\simeq 1.25$ \cite{Leschhorn}.

\section{Numerical results} 

\subsection{Constant velocity.} \label{sec:3}

The primary interest of this study is to compute the exponent $\tau$ which 
characterizes the collective motion of the particles at $F\simeq F_c$, 
and in particular its dependence on the driving 
velocity. We note that the avalanche
exponent was found to decrease with the driving velocity
in the Barkhausen effect, due to the motion of domain
walls in a ferromagnet \cite{Zapperi1}. The same effect was observed in the 
friction experiments reported in Ref~\cite{Laroche}.
 
\begin{figure}  
\epsfig{file=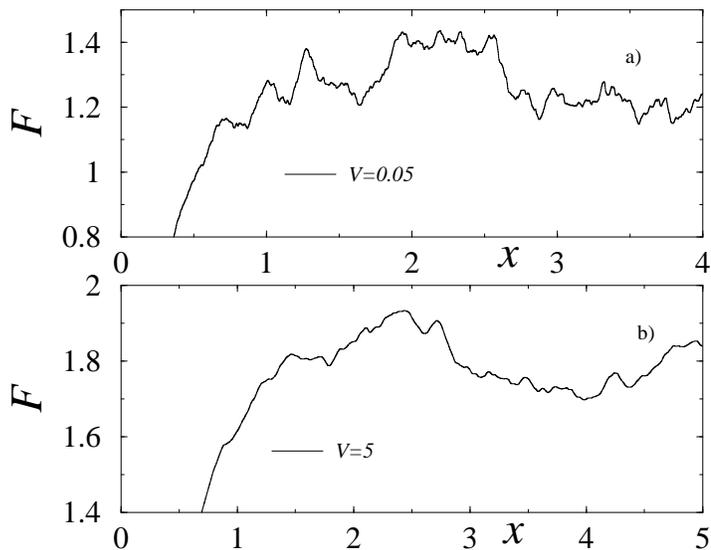,width=\linewidth} 
\caption{Friction force as a function of the displacement 
for different loading speeds. a) $V=0.05$, b) $V=5$.} 
\label{fig:courbe_vue}
\end{figure}

In order to reach the scaling regime, we progressively decrease 
$V$ and $K$ and compute the friction force.
Fig.~\ref{fig:courbe_vue} shows two typical plot for the friction force
as a function of the position of the line, in 
Fig.~\ref{fig:courbe_vue}a the driving velocity is 
$V=0.05$ and in Fig.~\ref{fig:courbe_vue}b $V=5$. As the driving velocity
increases the friction force becomes smoother, and in the limit 
$V \gg 1$ we obtain a viscous behavior ($F \sim \eta V$) 
with small relative fluctuations.
On the contrary,  for small velocities the dynamics is jerky:
the force increases with time until the  
beads are sufficiently stressed so that the chain depins 
decreasing the force. In this case, the friction force
displays a characteristic stick-slip pattern.

\begin{figure}  
\epsfig{file=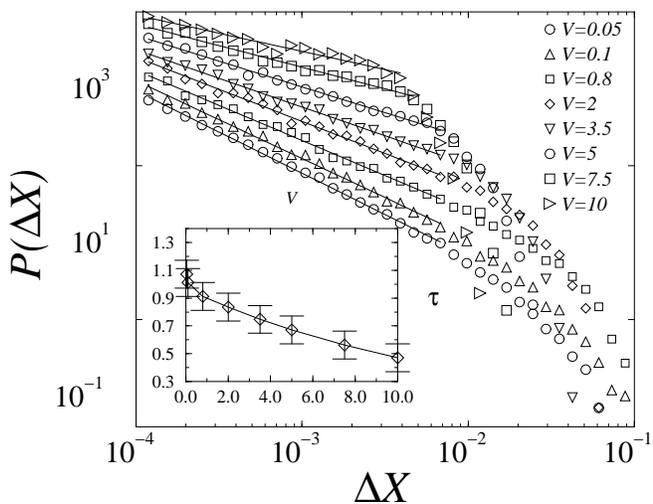,width=\linewidth} 
\caption{
Log-log plot of the size distribution 
of slip events for different values 
of $V$. In the inset we report the exponent $\tau$,
obtained fitting the distributions, as a function
of the driving velocity.}
\label{fig:courbe_tau_v}
\end{figure}

We thus measure the friction force drops $\Delta F$, or the
slip sizes $\Delta X= \Delta F/K$ and analyze their distribution.
The distributions are averaged over 20 realizations of the disorder
for $V=0.05$ and 100 realizations for $V=10$; 
in all cases the system was composed  of $L=1000$ beads, and the 
disorder was produced by $N=20000$ pining sites, Poissonian distributed.
The value of the exponent $\tau$ is obtained 
by a direct fit of the linear part of the distribution 
plotted in a log-log graph.  The results are shown in 
Fig.~\ref{fig:courbe_tau_v}, the main graph presents 
the log-log plot of the probability distribution function of the jumps 
for  various $V$ and in the inset we report the value of $\tau$ as  function 
of the velocity. We see a slow decrease of the exponent when the 
velocity increases. This result is in good qualitative agreement with the 
experiments reported in Ref.~\cite{Laroche}. 
The value of $\tau$ for $V\to 0$  agrees well with the exponent obtained in
elastic line depinning under quasistatic conditions \cite{Roux}
(see table I). 
   
\begin{figure} 
\epsfig{file=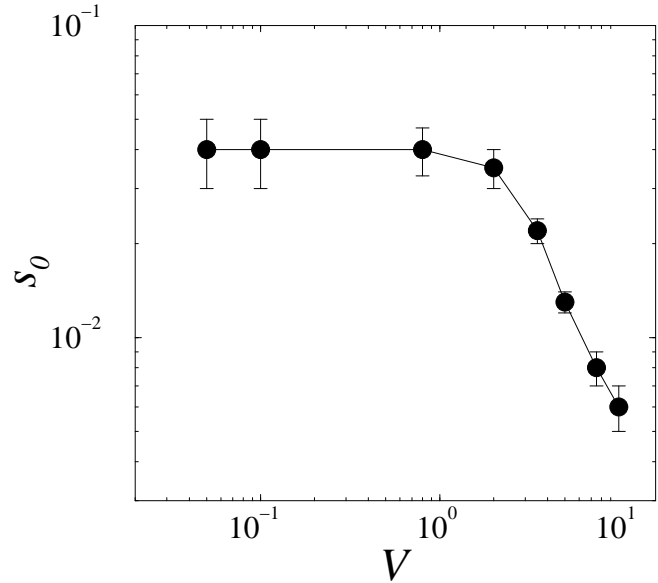,width=\linewidth} 
\caption{Cutoff of the slip size distribution 
as a function of $V$. For sufficiently low 
velocities ($V<V^*$) the cutoff is independent of $V$.} 
\label{fig:courbe_cut_V}
\end{figure}

\begin{figure} 
\epsfig{file=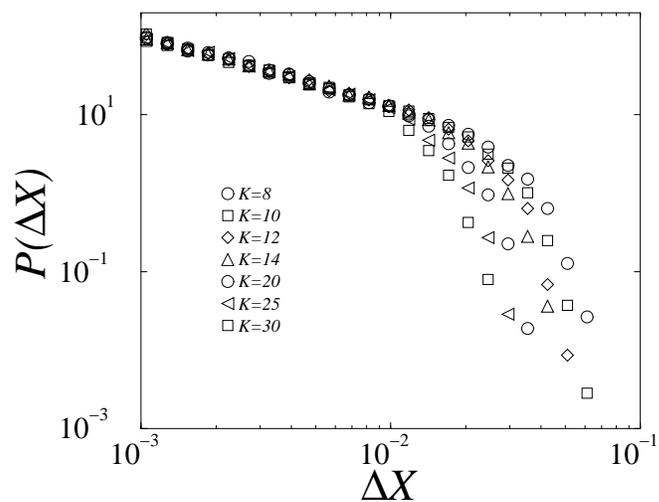,width=\linewidth} 
\caption{Distribution of the slips events as a function of $K$
showing that the cutoff  increases when $K$ decreases. 
The velocity $V$ is kept constant.}
\label{fig:courbe_tau_K}
\end{figure}

\begin{figure}  
\epsfig{file=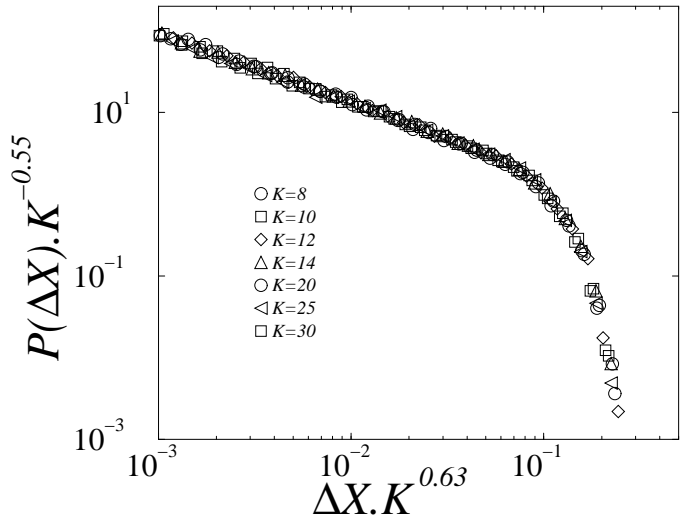,width=\linewidth} 
\caption{Data collapse of Fig.~\protect\ref{fig:courbe_tau_K}.}
\label{fig:courbe_tau_K_col}
\end{figure}

Fig.~\ref{fig:courbe_tau_v} shows that the cutoff $\Delta X_0$ 
clearly depends on $V$ when $K$ is hold fixed.  
We can quantify this variation and the result is 
reported in Fig.~\ref{fig:courbe_cut_V}. For small 
velocities ($V < V^*$) the cutoff is a constant ({\it{i.e}} 
it does not depend on $V$) which in principle depends on $K$,
while for high velocities the cutoff decreases with $V$, roughly
as a power law. Next, we study the behavior of the cutoff
when $V < V^*$ as $K$ is varied.
In Fig.~\ref{fig:courbe_tau_K} we shows the distribution 
of slip sizes for various $K$ for very small 
driving velocity. We see  that the cut off increases 
as $K$ is decreased. Fig.~\ref{fig:courbe_tau_K_col} shows the collapse 
of the curves after  the rescaling with $K$, in accordance with 
Eq.~(\ref{eq:scaling2}).

\begin{figure}  
\epsfig{file=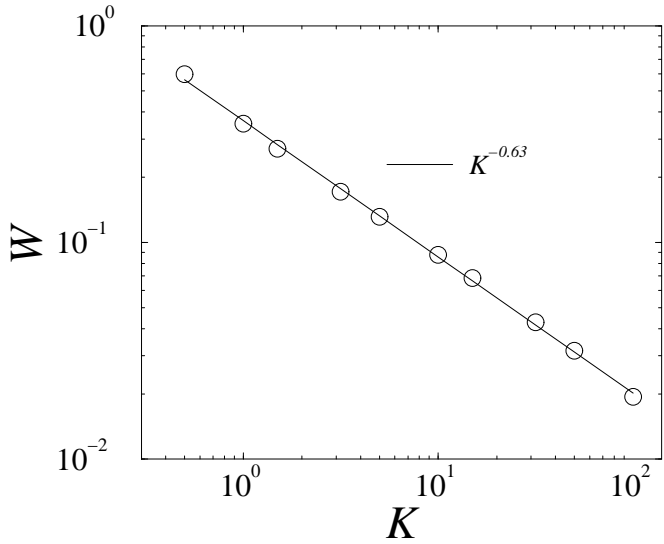,width=\linewidth} 
\caption{Width of the beads displacements
as a function of $K$. The power law fit of the 
curve and the scaling relations imply that $\zeta=1.26 \pm 0.03$.}
\label{fig:courbe_tau_rugo}
\end{figure}

We also measure the roughness exponent of the system following
Eq.~(\ref{eq:scaling_rugo}) and the result is reported in  
Fig.~\ref{fig:courbe_tau_rugo}. 
We obtain with both methods
$\zeta=1.26$, which is consistent with the numerical value found in
interface depinning \cite{Leschhorn},
for a disordered chain \cite{Hwa}, and with a  recent two loop 
renormalization group calculation \cite{Wiese}.
This value nevertheless disagrees with
previous results on a periodic chain \cite{Hwa}.

\begin{figure}  
\epsfig{file=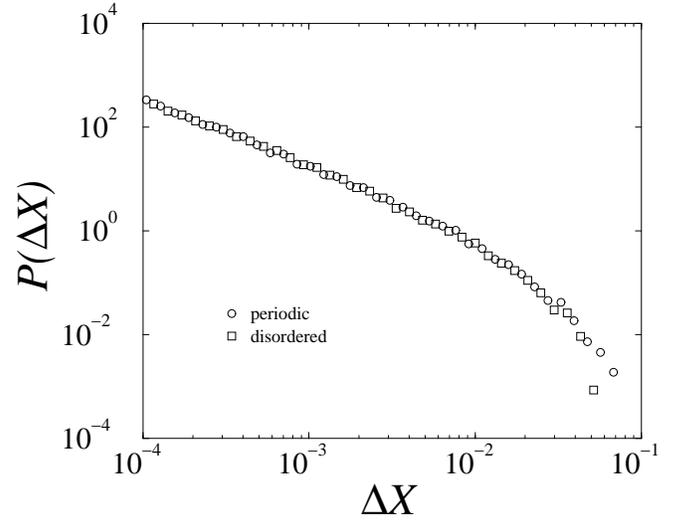,width=\linewidth} 
\caption{Probability distribution of slip sizes 
for periodic and disordered chain. We see no evidence for the 
existence of two different classes, since the distributions are
indistinguishable.} 
\label{fig:courbe_aleat_perio}
\end{figure}

The simulations of Ref.~\cite{Hwa}  for periodic and disordered chain,
suggest the presence of two different universality  classes. 
In order to test this result we study the force fluctuations
of a disordered chain. The equilibrium length of the springs
connecting the beads is chosen randomly (Poisson). 
The chain is then driven at constant velocity
and the distribution of the slip sizes is calculated. The result is shown in 
Fig.~\ref{fig:courbe_aleat_perio} where we also report 
the distribution obtained with a periodic chain using similar
parameters. The two distributions are clearly indistinguishable,
casting some doubt on the relevance of disorder in the spring
lengths. From this study, one would conclude that the two dynamics are in the 
same universality class.

\subsection{Constant force.} \label{sec:4}

For constant force driving, we employ system 
sizes varying from $L=20$ to $L=200$ and the density of 
pinning sites was chosen equal to unity $(L=N)$.
In order to determine the exponent $\beta$, we need an accurate
estimate of the critical force, since an error in $F_c$ can
strongly bias the fit.

\begin{figure}  
\epsfig{file=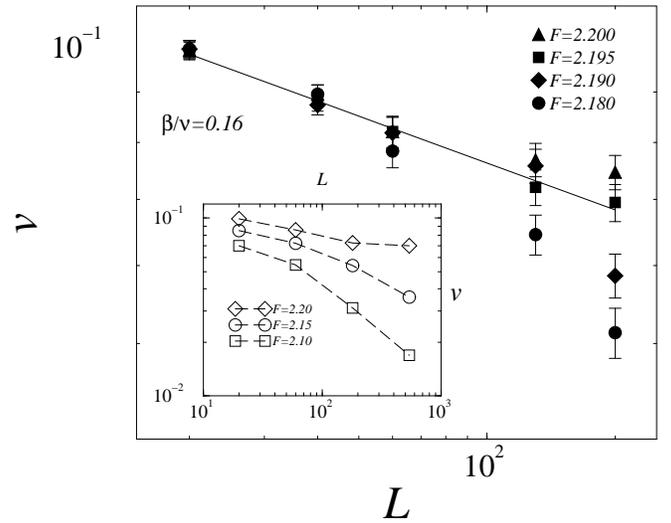,width=\linewidth} 
\caption{The variation of the mean velocity $v$ of the
chain as a function of the number of beads $L$ and for various forces
$F$. The inset shows the data for $F$ far from the critical force $F_c$,
and the main plot shows the data for $F$ near the critical force, 
$F_c \approx 2.195$.}
\label{fig:courbe_beta_nu}
\end{figure}

Fig.~\ref{fig:courbe_beta_nu} shows the value of the average velocity of the 
interface as a function of $L$. For $F$ smaller than 
$F_c$ in the limit of a large system $v$ tends to zero, and for $F$ greater 
than $F_c$ and for the same limit ($L \gg 1$) $v$ should tend a non
vanishing value. In this way we can locate the critical force, which results
to be   $F_c=2.195 \pm 0.005$. This result appears clearly from 
Fig.~\ref{fig:courbe_beta_nu}, from the log-log plot 
of $v$ as a function of $L$. We see that $F=2.195$ is compatible with a
power-law behavior, whereas for $F=2.200$ in the large
$L$ limit the mean velocity tends to a nonzero constant. For $F=2.190$  
the velocity tends to zero faster than a power law
in the limit of large system sizes. The numerical 
results are averaged over a number of
disorder configurations which varies from $4000$ for $L=20$ to $500$ 
for $L=200$.  The curve plotted in Fig.~\ref{fig:courbe_beta_nu} 
allows also to estimate $\beta / \nu = 0.16 \pm 0.02$
(see Eq.~(\ref{eq:scaling4})).

\begin{figure} 
\epsfig{file=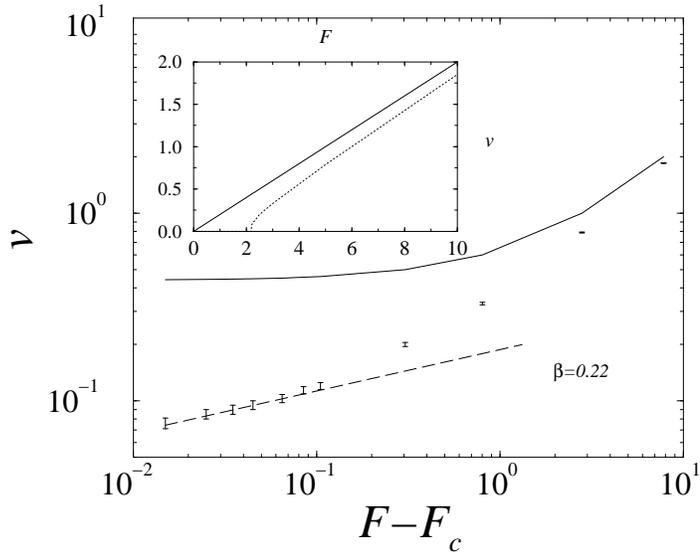,width=\linewidth} 
\caption{Average velocity $v$ as a function of the force $F$.
The inset shows the $v$-$F$ diagram. The solid curve represents 
the result without pinning centers and the 
dotted line is the result for the disordered substrate. The main graph
represents the log-log plot of $v$ as a function of $F-F_c$ 
The exponent $\beta$ is calculated using the six first 
points of the curve and the exponent is $\beta=0.22 \pm 0.02$.}
\label{fig:courbe_beta}
\end{figure}

Next, we calculate the exponent $\beta$ directly,
plotting $v$ vs $(F-F_c)$.
The fit in Fig.~\ref{fig:courbe_beta} yields $\beta=0.22 \pm 0.02$.
We restrict the fit to the the six smaller values since a crossover
to linear behavior is expected at high forces (i.e. $v=F/\eta$)
and this can bias the numerical estimate of the exponent. 
The simulations are made  with a system of $L=540$ particles, 
and the results are averaged over $100$
configurations of the disorder. In this way, we can obtain
$\beta$ and $\nu$ as summarized in table I.

\begin{figure}  
\epsfig{file=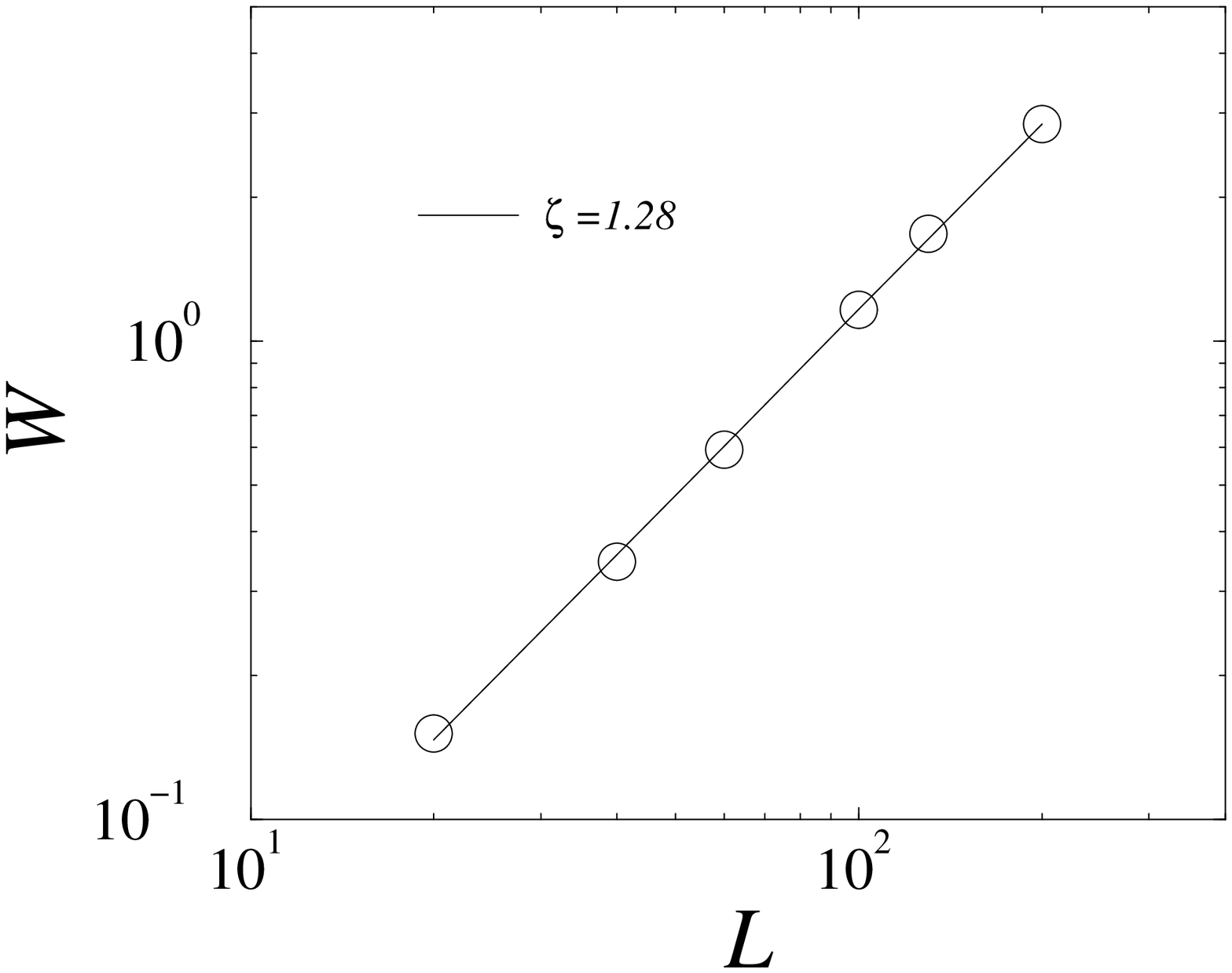,width=\linewidth} 
\caption{Width of the beads displacements $W$ as a function of $L$. The
scaling  relation between the two quantities gives $\zeta$. We recover  a
value  in good agreement with our previous results. }
\label{fig:courbe_beta_rugo}
\end{figure}

To further test the consistency of our results, we calculate the 
exponent $\zeta$ measuring the scaling of $W$ with $L$ at $F=F_c$
(see Eq.~\ref{eq:scaling5})
The results shown in Fig.~\ref{fig:courbe_beta_rugo} 
give $\zeta=1.28 \pm 0.03$ in agreement with the result obtained in 
Sec.~\ref{sec:3}.

\section{Conclusion}  \label{sec:conclusion}

In conclusion, we have investigated the dynamics of an elastic
chain sliding on a disordered substrate. We have analyzed 
numerically the scaling close to the depinning transition
focusing on the effect of different driving modes. 
Usually the problem is analyzed under constant force driving,
while friction experiments are usually performed controlling
the velocity. The two problems are closely related as 
discussed in Refs~\cite{Zapperi2,Hwa,tang-bak,maslov,Rev}. We compute the
critical exponents characterizing the transition and 
analyze the effect of the driving velocity and the
loading spring stiffness. Our results are in qualitative
agreement with friction experiments performed with 
macroscopic asperities coupled by an elastic matrix.
In general friction experiments our model will not
apply since in many instances the Larkin length is
extremely large \cite{Car-Noz}. In addition, inertial
effect are present in most cases and could lead to
different force fluctuations.

The exponents we measure agree well with the
values expected for the depinning  of elastic interfaces
in quenched disordered media \cite{Leschhorn} and with the
renormalization group calculation of Ref.~\cite{Wiese}. 
This is not 
surprising, since it is possible to show using the
method discussed in Ref.~\cite{Ledoussal} that 
the continuum limit of the model we study is described by 
\begin{equation}
\frac{\partial h(x,t)}{\partial t} = D \nabla h + F +f_p(x,h),
\end{equation}
where $h(x,t)$ is a coarse grained version of $u_i(t)$
and $f_p$ is a coarse grained random pinning force.
It is thus expected that the simulations performed in
Ref.~\cite{Hwa} for a periodic chain  agree with
this result. The only difference between those simulations
and ours lies in the way disorder is implemented:
in Ref.~\cite{Hwa} the pinning point are arranged in 
a periodic structure and have random strength, while
we use constant strength and random positions. We also 
note that the roughness exponent measured in  Ref.~\cite{Hwa}
is very close to $\zeta=3/2$, which is expected 
below the Larkin length, although the parameter employed
do not seem to be consistent with that regime. 

S. Z. acknowledges financial support from EC TMR Research Network
under contract ERBFMRXCT960062.


\begin{table}
\begin{tabular}{cc}
$\tau$   & 1.07 $\pm$ 0.05 \\
$\zeta$   & 1.26 $\pm$ 0.03\\
$\beta$   & 0.22 $\pm$ 0.02\\
$\nu$   & 1.3  $\pm$ 0.1\\
\end{tabular}
\caption{Critical exponents measured in simulations.}
\label{table}
\end{table}

\twocolumn 

\end{multicols}

\begin{thebibliography}{99}
\bibitem{note} another possibility is to drive the chain quasistatitcally
with an extremal dynamics \cite{maslov}.  The velocity is then constant
and equal to $1/L$ and the chain approches the critical point.
\end{thebibliography}
\end{document}